\documentclass[journal,onecolumn]{IEEEtran}
\ifCLASSINFOpdf
\else
\fi

\usepackage{graphicx}

\usepackage{tikz}
\usepackage{comment}
\usepackage{amsmath,amssymb} 
\usepackage{amssymb}
\usepackage{bbm}
\usepackage{algorithmic}
\usepackage{algorithm}
\usepackage{array}
\usepackage{color}

\usepackage{textcomp}
\usepackage{stfloats}
\usepackage{url}
\usepackage{verbatim}
\usepackage{cite}

\usepackage[accsupp]{axessibility}  

\usepackage[caption=false, font=footnotesize, labelfont=sf, textfont=sf]{subfig}

\usepackage[width=122mm,left=12mm,paperwidth=146mm,height=193mm,top=12mm,paperheight=217mm]{geometry}

\newcommand{\algrule}[1][.2pt]{\par\vskip.5\baselineskip\hrule height #1\par\vskip.5\baselineskip}
\newcolumntype{P}[1]{>{\centering\arraybackslash}p{#1}}

\begin{document}
%
\title{Edge-Aware Image Color Appearance and Difference Modeling}
%
%
%

\author{Abhinau K. Venkataramanan\\
\IEEEmembership{Department of ECE, UT Austin}}

\maketitle

\begin{abstract}
The perception of color is one of the most important aspects of human vision. From an evolutionary perspective, the accurate perception of color is crucial to distinguishing friend from foe, and food from fatal poison. As a result, humans have developed a keen sense of color and are able to detect subtle differences in appearance, while also robustly identifying colors across illumination and viewing conditions. In this paper, we shall briefly review methods for adapting traditional color appearance and difference models to complex image stimuli, and propose mechanisms to improve their performance. In particular, we find that applying contrast sensitivity functions and local adaptation rules in an edge-aware manner improves image difference predictions.
\end{abstract}


%
\IEEEpeerreviewmaketitle

\section{Introduction}
\label{ref:introduction}

A common notion of the defining property of ``color'' is generally the wavelength of visible electromagnetic (EM) radiation. However, such a description of color lacks a fundamental element - the (human) observer. Indeed, color is a perceptual phenomenon that is related to, but not entirely specified by, physical properties of surfaces such as reflectance spectra. Rather, the properties of the human visual system (HVS) and its response to the illumination of the environment, the brightness of the color, the presence of other colors during viewing, etc. play a crucial role in how the color is perceived. Under appropriately chosen conditions, the ``same color'' (physically, in terms of reflectance/emission spectra) may be perceived as being different, or vice versa. 

\begin{figure}
    \centering
       \includegraphics[width=0.45\linewidth]{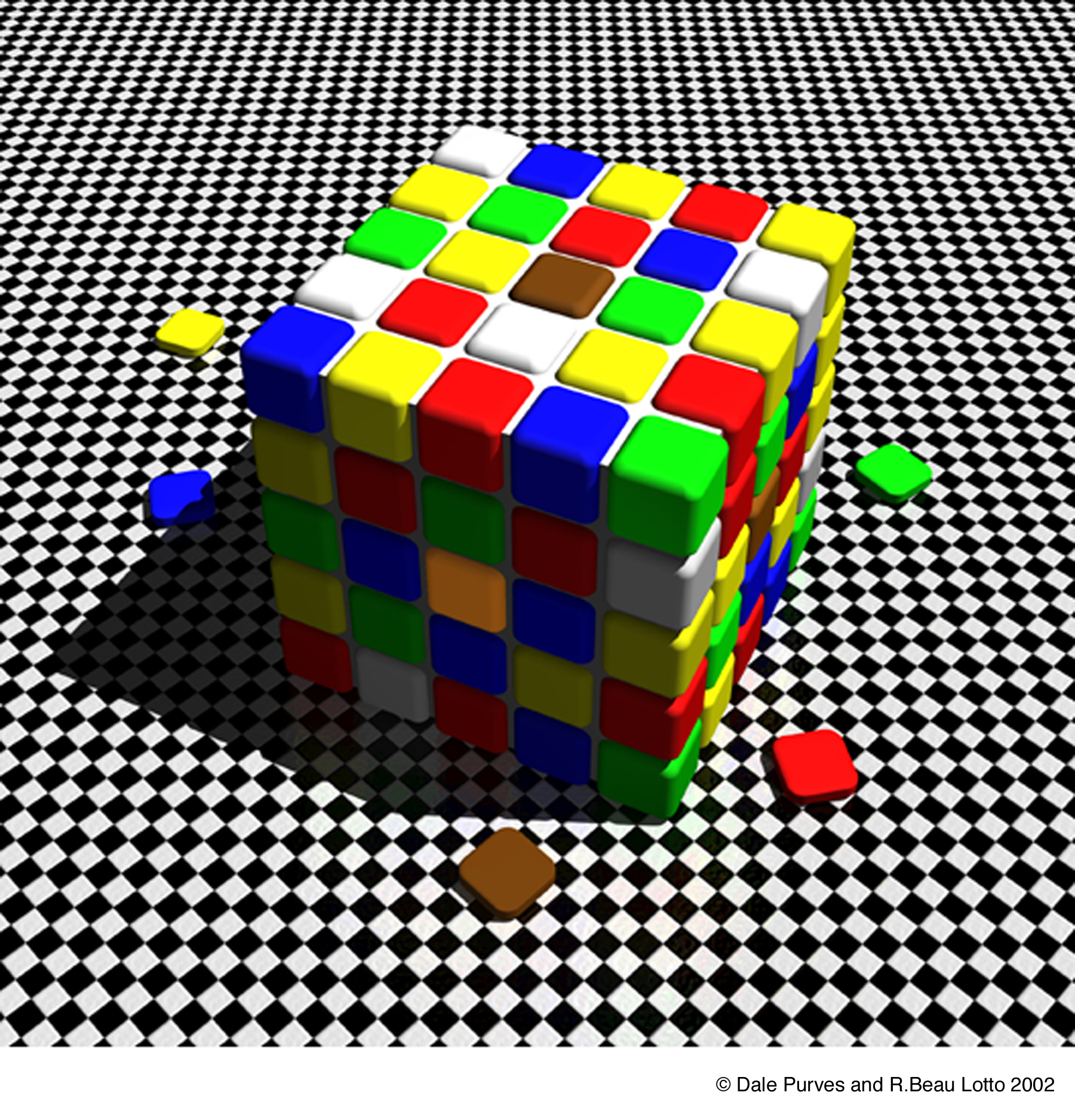}
    \caption{The color cube illusion}
    \label{fig:colorcube}
\end{figure}

A classic example of such a phenomenon is the color cube illusion \cite{ref:colorcube}, shown in Figure \ref{fig:colorcube}, in which the ``brown patch'' at the center of the top face and the ``orange patch'' at the center of the side face are identical, but appear to be different. This may be attributed to our perception of the front face of the cube being in a shadow. Furthermore, Lateral Inhibition (LI) leads to a reduction in the sensitivity of the HVS to low-frequencies \cite{ref:li_connections}, which gives rise to the characteristically band-pass contrast sensitivity function (CSF).

Therefore, when assessing differences in appearance between two images, it is important to account for local phenomena, both regarding adaptation to illumination and the sensitivity of the eye to spatial patterns of different frequencies. This is precisely the approach taken by Image Color Appearance \cite{ref:icam02} and image difference \cite{ref:idiff} models. In particular, these models apply local adaptation rules and incorporate models of the CSF to adapt traditional color appearance models, which have been developed for uniform color patches, to predict the appearance of, and differences between, local regions in images.

In this paper, we posit that human perceptions of appearance are made somewhat independently between (distinct regions) of objects. For example, when applying the CSF to account for the visibility of various artifacts, we aim to minimize the ``leakage'' of the influence of the CSF across object boundaries, i.e., edges. A similar approach is adopted when estimating local adaptation parameters, thereby reducing the influence of objects on each others' appearance. Using this, we demonstrate that such ``edge-aware'' image difference models better localize differences in the appearances of objects.

The rest of the paper is organized as follows. Section \ref{sec:color_diff_appearance} first presents an overview of the color difference modeling problem and Section \ref{sec:ucs} presents Uniform Color Spaces (UCSs), which are simple models designed to predict color differences. Section \ref{sec:cam} describes standard Color Appearance Models (CAMs), which build on UCSs to account for how the HVS adapts to illumination conditions, and Section \ref{sec:csfs} describes models of contrast sensitivity, which are used to account for the change in visibility with spatial frequencies.

Section \ref{sec:image_diff} describes the iCAM02 \cite{ref:icam02} and iDiff \cite{ref:idiff} models, which are the image color appearance and difference models that form the basis of this work. Section \ref{sec:edge_aware} describes the proposed edge-aware iCAM and iDiff frameworks, and Section \ref{sec:experiments} describes the experiments used to validate our method. Finally, we summarize the work presented in this paper and point out areas for future work in Section \ref{sec:conclusion}.

\section{Color Difference Modeling}
\label{sec:color_diff_appearance}
The simplest question that one can ask regarding the appearance of colors is whether two colors appear identical. Such questions are answered using psychophysical experiments, where subjects may be shown two colors and their responses recorded. Psychometric functions are then used to build models of an ``average'' observer, and these functions may depend on various factors such as the color of the ``reference'' stimulus, the luminance to which the eye has adapted, the background against which the stimulus was viewed, and in some cases, the spatial frequency characteristics of the stimulus.

For instance, such experiments were carried out by \cite{ref:scielab,ref:colorfest,ref:sccsf} to develop models of ``chromatic contrast sensitivity''. In particular, \cite{ref:colorfest} and \cite{ref:sccsf} consider several ``chromatic-directions'', such as varying colors from red to green, yellow-green to violet, etc. A detailed description of contrast sensitivity functions is given in Section \ref{sec:csfs}

Given that we are able to predict whether two colors appear identical to an average observer, under given viewing conditions, a more sophisticated task would then be to quantify the difference between the colors. Such models are termed as ``color-difference'' models. Finally, one may also desire ``color appearance models'' (CAMs) that directly predict quantities that correlate with human perception, called ``appearance correlates'', such as brightness, colorfulness, hue, etc. Note that while color-difference models typically use color appearance models to predict appearance correlates and measure their difference, this is not a necessary property of color-difference models. For instance, the \(\Delta E_{ab}\) metric computed using the CIELAB color space, described in Section \ref{sec:ucs}, is computed from an opponent color space, and \cite{ref:sccsf} models a ``probability of detection'' function that quantifies the difference between two colors directly from an LMS color space \cite{ref:cie_lms} that aims to model cone responses.

\section{Uniform Color Spaces}
\label{sec:ucs}

While color representations such as \(XYZ\) offer a device-independent representation of color, and \(LMS\) values characterize cone responses, neither representation can be directly used to predict differences between colors. In particular, (Euclidean) distances between two colors in the \(XYZ\) and \(LMS\) spaces are not ``perceptually uniform'', i.e., equal distances between points do not correspond to equal perceived differences in color.

\begin{figure}[ht]
    \centering
    \includegraphics[width=0.5\linewidth]{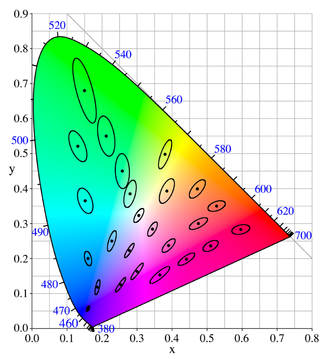}
    \caption{MacAdam Ellipses on the \(xy\) Chromaticity Plane}
    \label{fig:xy_macadam}
\end{figure}

This phenomenon is visualized in Figure \ref{fig:xy_macadam}, which shows the \(xy\) chromaticity plane with MacAdam's ellipses \cite{ref:macadam} overlaid on top. Each ellipse represents the set of colors that are ``just-noticeably-different'' from the color at the center. Note that the ellipses have been scaled to 10x their size for visibility. The green region at the top occupies a large region of slowly varying colors, while colors change much quicker in the red and yellow regions at the bottom-right. 

A Uniform Color Space (UCS) is one in which Euclidean distances between colors predict the perceptual difference between them. In other words, two pairs of colors that have the same Euclidean distance between them appear equally different. The CIELAB model is arguably the most popular UCS in use today. CIELAB transforms input tristimulus \(XYZ\) values to an opponent-color space \cite{ref:hering} using information regarding the achromatic white reference illuminant, also described using tristimulus values as \(\textit{XYZ}_n\). The transformation is carried out by applying a ``compressive nonlinearity'' \(f\) to the ratios \(X/X_n\), \(Y/Y_n\), and \(Z/Z_n\), followed by transformation to an opponent color space
\begin{align}
    L^{*} &= 116f\left(\frac{Y}{Y_n}\right) - 16, \label{eq:lab1}\\
    a^{*} &= 500\left(f\left(\frac{X}{X_n}\right) - f\left(\frac{Y}{Y_n}\right)\right), \\
    b^{*} &= 200\left(f\left(\frac{Y}{Y_n}\right) - f\left(\frac{Z}{Z_n}\right)\right), \label{eq:lab3}
\end{align}
where
\begin{equation}
    f(x) = \begin{cases}
                \sqrt[3]{x} & \text{if } t > \delta^3 \\
                \frac{x}{3\delta^2} + \frac{4}{29} & \text{else}
           \end{cases},
\end{equation}
and \(\delta = 6/29\).

CIELAB has proven to be a popular UCS, particularly due to the widespread adoption of the \(\Delta E_{00}\) color difference metric \cite{ref:delta_e_2000}. \(\Delta E_{00}\) improves upon a previous version, called \(\Delta E_{94}\), and the simple Euclidean distance 
\begin{equation}
    \Delta E_{ab} = \sqrt{(L_1^{*} - L_2^{*})^2 + (a_1^{*} - a_2^{*})^2 + (b_1^{*} - b_2^{*})^2},
    \label{eq:delta_e_ab}
\end{equation}
which is the most straightforward application of the UCS construction.

A cylindrical transformation is typically applied to opponent color representations to derive other attributes of color. A cylindrical transformation applied to the CIELAB color space yields the CIELCh color space, where
\begin{equation}
    C^{*} = \sqrt{a^{*2} + b^{*2}}
\end{equation}
denotes the ``chroma'' or ``relative-saturation'', and 
\begin{equation}
    h = \arctan\left(b^{*} / a^{*}\right)
\end{equation}
denotes the ``hue angle''. When comparing two colors, in addition to the total difference \(\Delta E_{00}\), these quantities can be used to predict differences in lightness, chroma, and hue \cite{ref:hue_diff}, as
\begin{equation}
    \Delta L^{*} = L_1^{*} - L_2^{*}
    \label{eq:l_diff}
\end{equation}
\begin{equation}
    \Delta C^{*} = C_1^{*} - C_2^{*}
    \label{eq:c_diff}
\end{equation}
\begin{equation}
    \Delta H = 2\sqrt{C_1^{*} C_2^{*}}\sin\left(\frac{h_1 - h_2}{2}\right)
    \label{eq:h_diff}
\end{equation}

Modern UCS models modify the CIELAB framework by first transforming input \(XYZ\) tristimulus values to an \(LMS\) space that predicts cone responses. Then, a compressive non-linearity is applied to these LMS responses, and a transformation is carried out to an opponent color space. IPT \cite{ref:ipt} and OKLab \cite{ref:oklab} are two simple modern UCS models that follow this framework. IPT improves on CIELAB in terms of its hue prediction, while OKLab improves over IPT in terms of its lightness and chroma predictions.

These approaches have also been extended to high-dynamic range (HDR) colors that cover a greater range of luminances and a wider gamut of chromaticities. The typical approach remains the same, with modifications made to the compressive non-linearities (HDR-CIELAB and HDR-IPT \cite{ref:hdr_cielab_ipt}), or a recalibration of the entire model (ICtCp \cite{ref:itu_bt2100} and \(\text{J}_\text{z}\text{a}_\text{z}\text{b}_\text{z}\) \cite{ref:hdr_ucs}) while following a similar sequence of operations.

\section{Color Appearance Models}
\label{sec:cam}
The CIELAB uniform color space \cite{ref:cielab}, while not being a ``true color appearance model'', is worth discussing in this context since it is a precursor to current widely-used CAMs. A key weakness of the CIELAB space is ``color inconstancy''. That is, CIELAB difference values are poor predictors of color constancy under varying illuminations \cite{ref:cielab_color_constancy}. This behavior is attributed to the ``improper'' method used by CIELAB to account for illumination conditions.

Due to the normalization of \(XYZ\) values against the reference white, the CIELAB transform makes a simple attempt at accounting for the appearance of the same color under various illumination conditions, hence its inclusion as a precursor to modern CAMs. However, such a normalization is considered an ``incorrect'' model of adaptation to reference whites, and has been shown to be a significant source of the color constancy problem described above \cite{ref:cielab_color_constancy}.

Rather, the ``correct'' method to account for the adaptation of the human eye to varying illuminations is von-Kries adaptation \cite{ref:von_kries}. von-Kries adaptation is based on the von-Kries coefficient rule, which states that for any observer, the cone responses to the same stimulus viewed under two different illuminations are proportional to the cone responses for the reference white under the two illuminations. 

In other words, let \(\textit{LMS}_1\), \(\textit{LMS}_2\) denote the cone responses for the same stimulus under two different illuminations. Furthermore, let \(\textit{LMS}_{w1}\) and \(\textit{LMS}_{w2}\) denote the cone responses for the reference white under the same two illuminations. Then,
\begin{equation}
    \begin{bmatrix}
        L_2 \\
        M_2 \\
        S_2 \\
    \end{bmatrix} = \begin{bmatrix}
        L_{w2}/L_{w1} & 0 & 0 \\
        0 & M_{w2}/M_{w1} & 0\\
        0 & 0 & S_{w2}/S_{w1} \\
    \end{bmatrix} \begin{bmatrix}
        L_1 \\
        M_1 \\
        S_1 \\
    \end{bmatrix}
\end{equation}

This equation, which predicts cone responses under varying illumination conditions is called von-Kries adaptation. In general, similar ``diagonal'' adaptation rules in the LMS color space are called generalized von-Kries transforms \cite{ref:generalized_von_kries}, and are considered the ``correct'' method for performing chromatic adaptation. In fact, a modification of the CIELAB UCS, called Conelab2, that performs von-Kries adaptation in the LMS space was found to exhibit better color constancy \cite{ref:conelab}.

The success of CIELAB led to the first standardized color appearance model published by CIE - CIECAM97s \cite{ref:ciecam97s}. The general approach taken by CIECAM97s can be described as follows, and this has been used as a template for modern CAMs.

\begin{enumerate}
    \item Convert the input \(XYZ\) tristimulus values to an \(LMS\) or ``sharpened'' \(LMS\) color-space. ``Sharpened'' \(LMS\) spaces are those that do not accurately reflect cone responses but achieve better appearance prediction.
    \item Calculate, or set, a degree of adaptation \(D\) based on the viewing conditions.
    \item Apply a von-Kries transform to adapt input colors to a reference illuminant, typically the equal energy illuminant \(XYZ = (100, 100, 100)\).
    \item Compress adapted \(LMS\) values using a generalized Michaelis-Menten equation \cite{ref:michaelis_menten}.
    \item Transform compressed \(LMS\) responses to an opponent color space.
\end{enumerate}

Steps 1-3 together constitute the Chromatic Adaptation Transform (CAT) part of the CAM, and CIECAM97s uses the Bradford CAM \cite{ref:bradford}. CIECAM97s was followed by a simpler and improved model called CIECAM02 \cite{ref:ciecam02}, which uses the CAT02 model of chromatic adaptation. CAT02 is a modified version of the CMCCAT2000 \cite{ref:cmccat2000} model, and was proposed as a simpler linear alternative \cite{ref:ciecam97s_improvement} to the Bradford CAT used in CIECAM97s. As a result, CAT02 and CIECAM02 have emerged as, by far, the most popular CAT and CAM models in use today. The CAT16 model attempts to improve upon CAT02 by simplifying it further into a one-step transform, and the corresponding CAM is called CAM16 \cite{ref:cam16}.

\section{Models of Contrast Sensitivity}
\label{sec:csfs}
The color difference models described in the aforementioned settings have all been developed based on psycho-visual experiments that recorded visible differences between patches of colors. However, images are not ``flat'' or ``constant-value'' patches. Rather, images are comprised of spatially varying stimuli, which are most easily decomposed into spatial sine waves of varying frequencies using the Fourier Transform. 

Sine wave gratings have been considered a good method for analyzing the visual systems, since they mirror the typical analysis of linear systems, and they allow for the use of Fourier Analysis to understand the visibility of more complex stimuli \cite{ref:csf2}. The simplest form of the experiment presents subjects with achromatic sine wave gratings or Gabor patches having a given spatial frequency, and subjects are allowed to change the contrast of each stimulus until it is ``just noticeable'' \cite{ref:csf1,ref:csf2}. The minimum visible contrast is called the detection threshold, and in its simplest form, is expressed as a function of the spatial frequency \(thresh(f)\). Then, the ``contrast sensitivity function'' (CSF) is defined as
\begin{equation}
    CSF(f) = 1/thresh(f)
\end{equation}

This analysis has been extended to opponent color spaces to also measure chromatic CSFs in the red-green and blue-yellow channels \cite{ref:opp_csf}, and along a wider set of ``chromatic directions'' \cite{ref:chromcsf_1,ref:chromcsf_2}. The data obtained from such experiments has been used to develop mathematical models of contrast sensitivity \cite{ref:mannos,ref:movshon}. Another approach to measuring CSFs involves recording the detectability of noise in stimuli, and this has been used to develop CSF models of both sine wave gratings \cite{ref:johnson} and wavelet subbands \cite{ref:watson}. Modern CSF models also account for the background luminance, angular size of the object, models of the optical system, etc \cite{ref:sccsf,ref:barten}.

\section{Image Difference Modeling}
\label{sec:image_diff}
While CSF models are more complex than simple CAMs due to the inclusion of spatial frequency, sine-wave gratings and Gabor patches are still considered simple stimuli compared to real-world images, since images exhibit a complex interaction of spatially-varying stimuli. Furthermore, CSF models are not directly equipped to predict either color appearance or differences. Therefore, there is a need for combining visual adaptation and contrast sensitivity effects with color appearance/difference models, particularly for use in the context of image difference prediction.

The image Color Appearance Modeling (iCAM) framework was proposed by Fairchild et al. in \cite{ref:icam02} that sought to adapt traditional CAM models to images and built on previous approaches such as the Spatial CIELAB (S-CIELAB) \cite{ref:s_cielab} and Visible Differences Predictor (VDP) \cite{ref:vdp} models. In short, the iCAM framework takes an image as input and transforms it to the IPT UCS, after applying local adaptation rules. The IPT values may be used to predict appearance correlates. A detailed description of iCAM02 is provided in Algorithm \ref{alg:icam02}. An updated version of the model, named iCAM06 \cite{ref:icam06} has also been proposed. However, since iCAM06 is focused on rendering HDR images, we do not include it in our discussion.

\begin{algorithm}
    \caption{The iCAM02 Image Color Appearance Model}\label{alg:icam02}
    \textbf{Input:} \\
    \hspace*{\algorithmicindent} \(XYZ(i, j)\) - Input image in CIE XYZ values.\\
    \textbf{Output:} \\
    \hspace*{\algorithmicindent} \(\textit{IPT}_a(i, j)\) - IPT values that account for local adaptation.
    \algrule
    \begin{algorithmic}
    \STATE \begin{itemize}
        \item Compute the local adaptation white point.
        \[XYZ_{white} \gets GaussianFilt(XYZ)\]
    \end{itemize}
    
    \STATE \begin{itemize}
        \item Use CAT02 to adapt \(XYZ\) values to the D65 illuminant.
        \[XYZ_{a} \gets CAT02(XYZ, XYZ_{white}, D65)\]
    \end{itemize} 

    \STATE \begin{itemize}
        \item Convert adapted \(XYZ\) values to an \(LMS\) space.
        \[LMS_{a} \gets M_{XYZ\rightarrow LMS} \times XYZ_{a}\]
    \end{itemize}

    \STATE \begin{itemize}
        \item Compute the luminance-level adaptation factor, normalized to have value 1 at 1000 nits.
        \[k \gets 1 / \left(1 + Y_{white}\right)\]
        \[F_L \gets \left((1/5) k^4 Y_{white} + (1/10) \left(1 - k^4\right)^2 Y_{white}^{1/3}\right)/1.71\]
    \end{itemize} 
    
    \STATE \begin{itemize}
        \item Use \(F_L\) to modulate the exponent of the IPT compressive nonlinearity.
        \[\alpha \gets 0.43 \max\left(F_L, 0.3\right)\]
    \end{itemize} 
    
    \STATE \begin{itemize}
        \item Apply the modulated compressive non-linearity.
        \[LMS_{a}^{'} \gets \text{sign}\left(LMS_{a}\right) \mid LMS_a\mid^{\alpha}\]
    \end{itemize}
    
    \STATE \begin{itemize}
        \item Convert to the opponent IPT UCS.
        \[IPT_{a} \gets M_{LMS\rightarrow IPT} \times LMS_{a}^{'}\]
    \end{itemize}
    \end{algorithmic}
\end{algorithm}
A simple extension of iCAM02 to predict image differences is to compute \(\Delta E\) metrics from the predicted IPT outputs when iCAM02 is applied to the two images. We shall overload notation and refer to this metric also as iCAM02, since we shall only discuss image difference models in this paper. A more sophisticated approach to predicting image differences may be constructed by combining the iCAM02 framework with the modular image difference (iDiff) \cite{ref:idiff} framework. A detailed description of iDiff is provided in Algorithm \ref{alg:idiff}.

\begin{algorithm}
    \caption{The iDiff Modular Image Difference Algorithm}\label{alg:idiff}
    \textbf{Input:} \\
    \hspace*{\algorithmicindent} \(\text{XYZ}_1(i, j), \text{XYZ}_2(i, j)\) - Input images in CIE XYZ values.\\
    \textbf{Output:} \\
    \hspace*{\algorithmicindent} \(\Delta E(i, j)\) - Total Error map. \\
    \hspace*{\algorithmicindent} \(\Delta I(i, j)\) - Brightness/Lightness difference map. \\
    \hspace*{\algorithmicindent} \(\Delta C(i, j)\) - Chroma difference map. \\
    \hspace*{\algorithmicindent} \(\Delta H(i, j)\) - Hue difference map.
    \algrule
    \begin{algorithmic}
        \STATE \begin{itemize}
            \item Convert both inputs into an opponent color space \cite{ref:wandell}
            \[\textit{ACC} \gets M_{XYZ\rightarrow ACC}\textit{XYZ}\]
        \end{itemize}
        \STATE \begin{itemize}
            \item Apply the NSS-adapted CSF to all three channels of both inputs
            \[\widetilde{\textit{ACC}} \gets \mathcal{F}^{-1}\left\{\mathcal{F}\left\{\textit{ACC}\right\} \times Flatten\left\{f_\theta^{1/3} MovshonCSF(f_\theta)\right\} \right\},\]
            where \(f_r = \sqrt{f_x^2 + f_y^2}\) is the radial frequency, \(\theta = \arctan\left(f_y/f_x\right)\) is the orientation, and \(f_\theta = f_r / \left(0.15 \cos(4\theta) + 0.85\right)\) accounts for the ``oblique effect'' \cite{ref:vdp}.
        \end{itemize}
        \STATE \begin{itemize}
            \item Preserve edges by performing ``edge-enhancement'' using the filter
            \[EdgeEnh(f_\theta) \gets 1 + e^{-(f_\theta - 30)^2/36}\]
        \end{itemize}
        \STATE \begin{itemize}
            \item Apply local contrast enhancement using gamma curves.
            \[Mask \gets GaussianFilter\left(\widetilde{A}\right)\]
            \[\beta \gets \text{clip}\left(2^{\frac{\text{median}\left(\widetilde{A}\right) - Mask}{\text{median}\left(\widetilde{A}\right)}}, -10, 10\right)\]
            \[\textit{ACC}^{'} \gets \left(\max_{ij}\{ACC\} - \min_{ij}\{ACC\}\right)\left(\frac{ACC}{\max_{ij}\{ACC\} - \min_{ij}\{ACC\}}\right)^{\beta}\]
        \end{itemize}
        \STATE \begin{itemize}
            \item Convert back to XYZ.
            \[\textit{XYZ}^{'} \gets M_{ACC\rightarrow XYZ}\textit{ACC}^{'}\]
        \end{itemize}
        \STATE \begin{itemize}
            \item Convert to IPT and compute differences as in Equations \ref{eq:delta_e_ab} and \ref{eq:l_diff}-\ref{eq:h_diff}.
        \end{itemize}
    \end{algorithmic}
\end{algorithm}

The iDiff framework makes use of CSF models to enhance or suppress frequencies, and therefore artifacts, based on their visibility. This is achieved by filtering both input images with CSFs by multiplying in the frequency domain. Curiously, the description of the CSF models used in iDiff is inconsistent with the constraints that the authors wish to enforce on them, and the plots showing the CSFs as a function of frequency. In particular, the authors claim to normalize CSFs to have unit DC values. However, the functional form of the ``base CSF'' prior to normalization is specified as 
\begin{equation}
     MovshonCSF(f) = af^ce^{-bf}.
\end{equation} 

Clearly, the response of this CSF at \(f = 0\) is 0, so it is unclear how normalizing by any divisive factor can yield a non-zero DC response, such as that in Figure 5 of \cite{ref:idiff_csf}. A similar issue is faced by CSFs that have been adapted to account for natural scene statistics, which involves multiplying ``base'' CSFs by a factor of \(f^{1/3}\).

To resolve this ambiguity, we propose using the following flattening rule proposed by \cite{ref:flattening} to convert bandpass CSFs into lowpass CSFs, which is said to account for supra-threshold effects.
\begin{equation}
    Flatten\left\{CSF(f)\right\} = \begin{cases}
        CSF(f_{max}) & \text{if }f \leq f_{max}, \\
        CSF(f) & \text{else},
    \end{cases}
\end{equation}
where \(f_{max} = \arg \max CSF(f)\). Therefore, all achromatic CSFs used in this work have been modified using the flattening method, and chromatic CSFs have been left unchanged since they are already low-pass. Furthermore, both classes of CSFs have been normalized to a peak response of 1.

Finally, we also consider the image difference model obtained by integrating iCAM02 and iDiff (say, iCAMDiff), which was proposed in \cite{ref:icam02}. This model involves applying the local adaptation rules of iCAM02 first, followed by the CSF-based difference model proposed by iDiff.

\section{Edge-Aware Filtering}
\label{sec:edge_aware}
The primary level of complexity that an image contains over a uniform or sine-wave patch is its spatial organization. Images contain structures at various scales, such as textures and objects, that are not easily captured by ``stationary'' models such as uniform and sine-wave patches. Indeed, even the use of Fourier analysis and filtering directly does not adequately capture the local spatial processing that occurs in the HVS. In particular, while the contrast sensitivity to high-frequency sine-wave gratings is low, edges are crucial visual cues that we use to ``segment'' images into objects and identify textures.

The edge enhancement applied in iDiff aims to correct the low CSF weighting applied to high frequencies by adding a Gaussian-shaped ``bump'' at high frequencies. However, it is important to note that edges are not high-frequencies, they merely contain them. The key property that defines edges is their spatial localization and non-stationarity, which is not accounted for by this model.

The key hypothesis of this paper is that the appearances of objects (and distinct regions of objects) in images are generally assessed separately from each other (though, not strictly independently). In other words, the ``leakage'' of color appearance and difference mechanisms across edges is to be minimized. To this end, we propose using edge-aware filtering, both to compute local adaptation white points in iCAM and to apply CSF models in iDiff. Such filtering may be interpreted as soft segmentation of regions that may correspond to distinct (regions of) objects.

The most popular edge-aware filter is the bilateral filter \cite{ref:bilateral}, which combines a Gaussian spatial weighting function \(w_s(\cdot)\), which is the ``base filter'', with a Gaussian ``influence function'' that depends on pixel intensities \(w_r(\cdot)\) as
\begin{equation}
    \widetilde{I}(i,j) = 
    \frac{\sum\limits_{(m,n) \in \Omega_{ij}} w_s\left(i-m, j-n\right) w_r\left(I\left(i,j\right) - I\left(m,n\right)\right)I\left(m,n\right)}{\sum\limits_{(m,n) \in \Omega_{ij}} w_s\left(i-m, j-n\right) w_r\left(I\left(i,j\right) - I\left(m,n\right)\right)}
    \label{eq:bilateral}
\end{equation}
where \(\Omega_{ij}\) denotes a spatial neighborhood centered on pixel location \((i,j)\).

The influence function assigns lower weights to neighboring pixels that differ significantly from the center pixel, thereby reducing the influence across edges, which are characterized by abrupt changes in pixel values. Since local adaptation white points are computed using a Gaussian filter in iCAM02, the bilateral filter is a natural candidate as the edge-aware alternative. 

To propose an edge-aware CSF filtering method, we generalize the spatial weighting function in the bilateral filter to include any linear filter. That is, we would like to replace the \(w_s\) function by the CSF filter. However, the most accurate method for applying the CSF is by multiplying in the frequency domain, which is incompatible with the spatial domain definition in Equation \ref{eq:bilateral}. Furthermore, no direct multiplication in the frequency domain would be equivalent to the desired filter, since edge-aware filtering is non-linear.

To address this, we adopt an approach similar to the fast bilateral filter proposed in \cite{ref:durand}, which we modify to improve numerical stability. Let us begin with a simple assumption, that the set of luminance values is a discrete set \(\{l_1, \dots, l_K\}\). Then, Equation \ref{eq:bilateral} may be rewritten as
\begin{align}
    \widetilde{I}(i,j) = 
    \frac{\sum\limits_{l_k}\left(\sum\limits_{(m,n)} w_s\left(i-m, j-n\right) w_r\left(l_k - I\left(m,n\right)\right)I\left(m,n\right)\right)M_k\left(i,j\right)}{\sum\limits_{l_k}\left(\sum\limits_{(m,n)} w_s\left(i-m, j-n\right) w_r\left(l_k - I\left(m,n\right)\right)\right)M_k\left(i,j\right)}
\end{align}
In short,
\begin{align}
    \widetilde{I} = \frac{\sum\limits_{l_k}\left(w_s \otimes \left\{W_{rk} \times I\right\}\right)M_k}{\sum\limits_{l_k}\left(w_s \otimes W_{rk}\right)M_k}
    \label{eq:fast_bilateral}
\end{align}
where \(W_{rk}(i,j) = w_r\left(l_k - I(i,j)\right)I(i,j)\) is an ``influence map'', \(M_k = \mathbbm{1}\left\{I_p = l_k\right\}\) is the binary mask denoting whether the value of the center pixel is \(l_k\), and \(\otimes\) denotes convolution. In general, if the luminance can take continuous values in the range \([0, L]\), we can divide the range into \(K\) bins, each centered at \(l_k\), and replace the binary mask by an interpolation function. 

In this manner, we have converted the computation of the non-linear bilateral filter into a sequence of linear filters. We may now replace the convolution-based filtering in Equation \ref{eq:fast_bilateral} with frequency domain filtering using the CSF, to yield an edge-aware CSF filter. Note that the influence function is computed using the intensity channel for all three channels, since edges are best identified in the intensity/luminance domain.

\section{Experiments}
\label{sec:experiments}
In this section, we shall compare edge-aware iCAM, iDiff and iCAMDiff models to their edge-unaware counterparts. We perform this analysis in two ways - by visualizing error maps and by observing the behavior of the predicted ``aggregate error'' for known distortions. In both cases, we shall rely on visual inspection and qualitative evaluations, since quantitative subjective data is not available. Moreover, as discussed in \cite{ref:idiff}, using the mean value as an aggregated difference metric is not preferred, since subjects are biased towards regions of low quality. A detailed analysis of spatial aggregation methods conducted in \cite{ref:essim} arrived at a similar conclusion. Minkowski pooling was identified in \cite{ref:essim} as an improvement over mean pooling, since it assigns higher weight to regions of higher distortion. Therefore, we compute the aggregate difference value from error maps as
\begin{equation}
    \Delta E_{agg} = \left(\frac{1}{MN}\sum\limits_{i,j} \Delta E(i,j)^3\right)^{1/3}.
\end{equation}

When using the flattening scheme proposed in Section \ref{sec:image_diff}, it was found that the NSS adaptation using the \(f^{1/3}\) factor did not change any of the results significantly. Furthermore, we use the CAT16 chromatic adaptation transform \cite{ref:cam16} and the OKLab UCS \cite{ref:oklab} in the place of CAT02 and IPT, since they have been proposed as improvements over the two models respectively.

To evaluate the proposed models, we have used tone-mapped HDR images, since tone-mapping is known to introduce both luminance and color artifacts. We use the Durand TMO \cite{ref:durand} and the Reinhard TMO \cite{ref:reinhard02} to generate two sets of videos, with varying contrast and color desaturation parameters. Note that while the two parameters are specific to achromatic and chromatic properties respectively in theory, they tend to influence both aspects of image appearance in practice.

\begin{figure} 
    \centering
  \subfloat[Contrast-Varied Images\label{fig:contrast_baseline}]{%
       \includegraphics[width=0.45\linewidth]{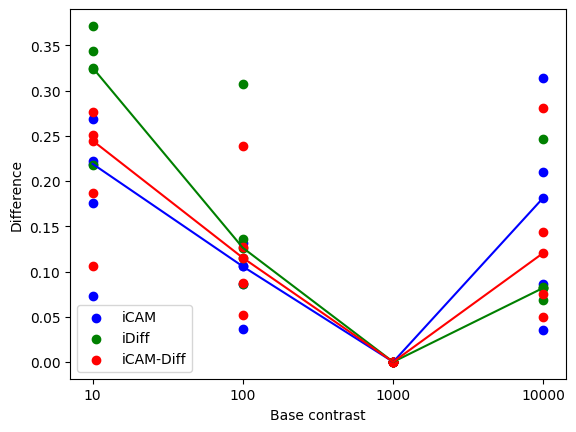}}
    \hfill
  \subfloat[Desaturation-Varied Images\label{fig:desat_baseline}]{%
        \includegraphics[width=0.45\linewidth]{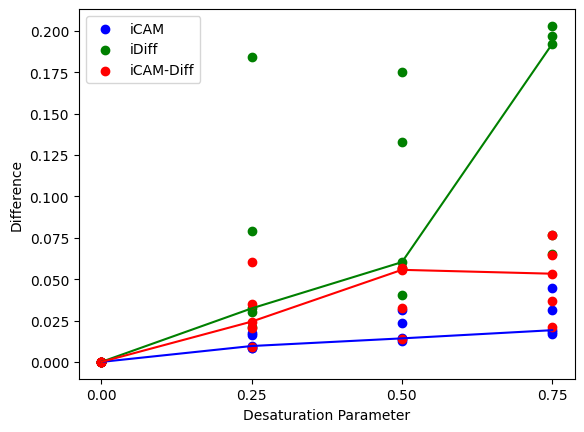}}
  \caption{Image difference predictions made by the baseline iCAM02, iDiff and ICAMDiff models}
  \label{fig:baseline_predictions} 
\end{figure}

\begin{figure} 
    \centering
  \subfloat[Contrast-Varied Images\label{fig:contrast_edge}]{%
       \includegraphics[width=0.45\linewidth]{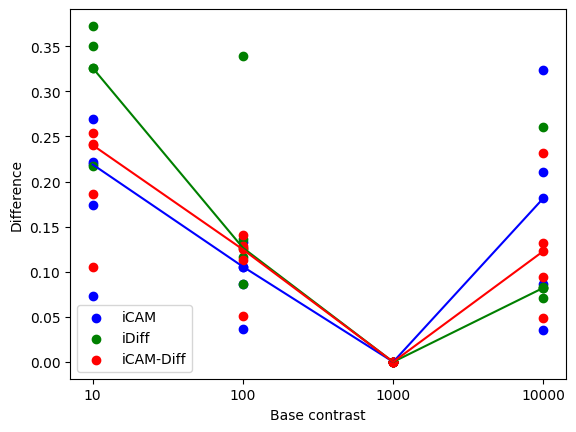}}
    \hfill
  \subfloat[Desaturation-Varied Images\label{fig:desat_edge}]{%
        \includegraphics[width=0.45\linewidth]{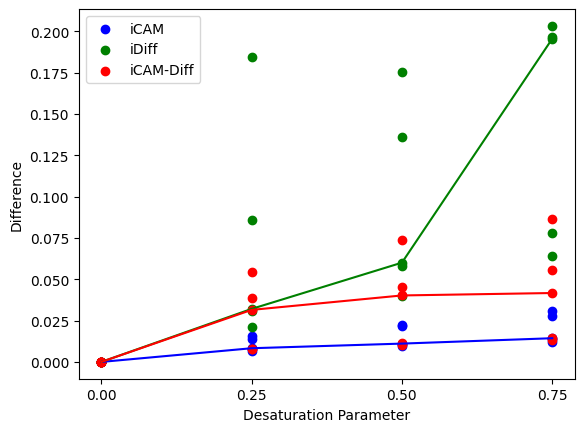}}
  \caption{Image difference predictions made by edge-aware iCAM02, iDiff and ICAMDiff models}
  \label{fig:edge_predictions} 
\end{figure}

The set of contrast-varied images were generated by setting the ``base contrast'' parameter to \(\{10, 100, 1000, 10000\}\). To compute the differences, images generated using a base contrast of 1000 were chosen as reference. The set of saturation-varied images were generated by setting the desaturation parameter to \(\{0.0, 0.25, 0.5, 0.75\}\), and the 0.0 desaturation images were used as references for difference calculation. For both sets of images, the aggregated differences predicted by iCAM02, iDiff, and iCAMDiff are shown in Figure \ref{fig:baseline_predictions}, and the predictions made by the proposed edge-aware counterparts are shown in Figure \ref{fig:edge_predictions}. 

From these figures, we observe that the edge-aware iCAMDiff model improves upon the baseline iCAMDiff model, which is the method recommended in \cite{ref:icam02}, in two ways. Firstly, the variation in quality over the contrast-varied images was reduced significantly, signaling a greater confidence in the predictions. Secondly, the baseline iCAMDiff model exhibits anomalous behavior at high desaturation values, predicting a lower difference on average for a desaturation parameter of 0.75 than for 0.5. Applying the model in an edge-aware manner fixes the non-monotonicity issue, and reveals the asymptotic behavior of perceived difference with increase in desaturation.

\begin{figure}
    \centering
  \subfloat[Reference Image]{%
       \includegraphics[width=0.45\linewidth]{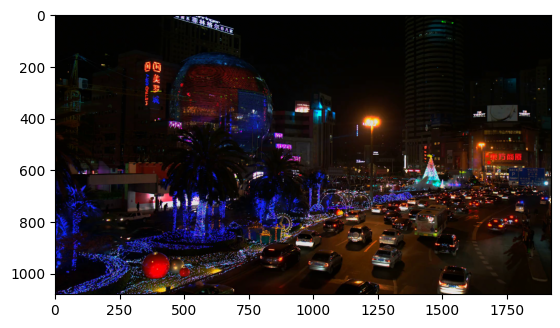}}
    \hfill
  \subfloat[Test Image]{%
        \includegraphics[width=0.45\linewidth]{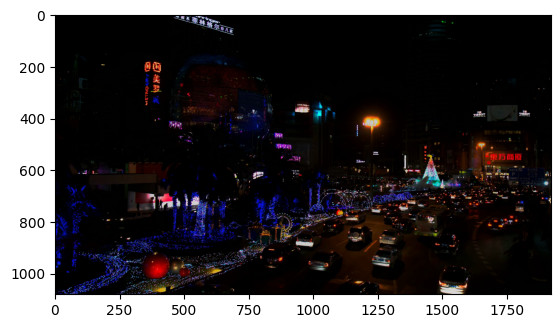}}
        \\
  \subfloat[Total Error (Baseline)]{%
       \includegraphics[width=0.45\linewidth]{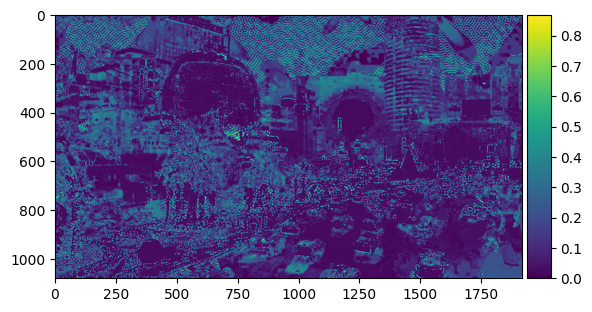}}
    \hfill
  \subfloat[Total Error (Edge-Aware)]{%
        \includegraphics[width=0.45\linewidth]{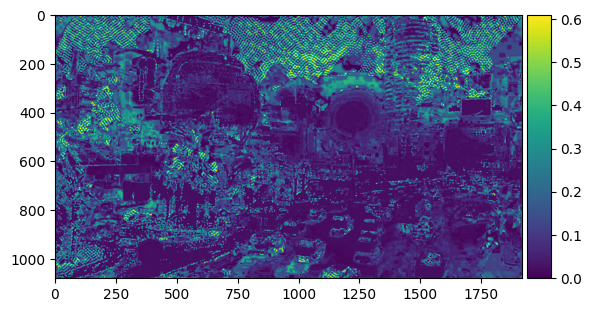}}
        \\
  \subfloat[Intensity Error (Baseline)]{%
       \includegraphics[width=0.45\linewidth]{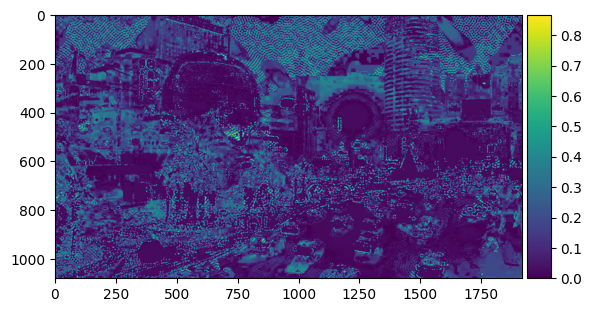}}
    \hfill
  \subfloat[Intensity Error (Edge-Aware)]{%
        \includegraphics[width=0.45\linewidth]{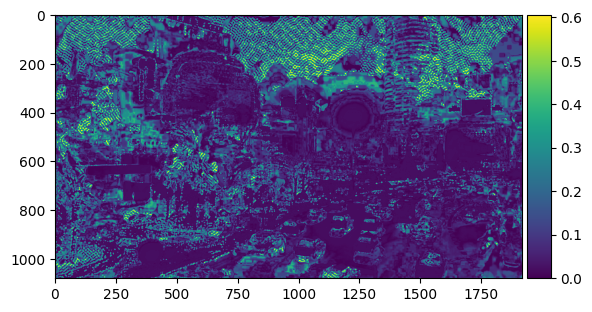}}
        \\
\subfloat[Chroma Error (Baseline)]{%
       \includegraphics[width=0.45\linewidth]{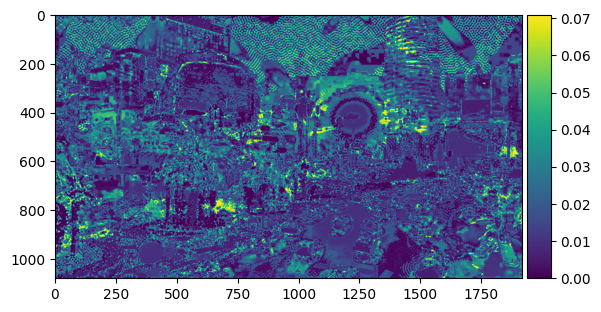}}
    \hfill
  \subfloat[Chroma Error (Edge-Aware)]{%
        \includegraphics[width=0.45\linewidth]{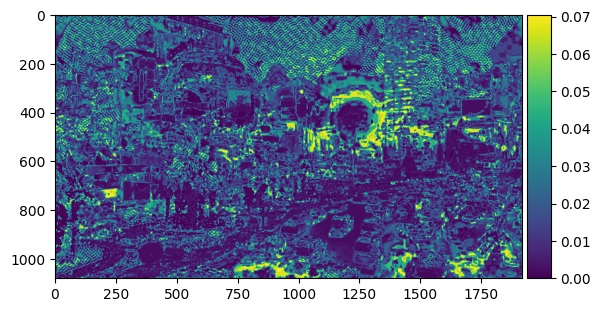}}
        \\
\subfloat[Hue Error (Baseline)]{%
       \includegraphics[width=0.45\linewidth]{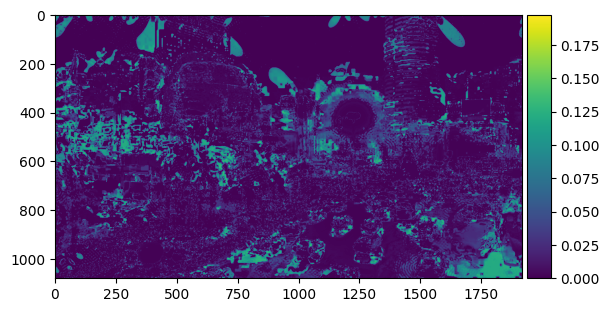}}
    \hfill
  \subfloat[Hue Error (Edge-Aware)]{%
        \includegraphics[width=0.45\linewidth]{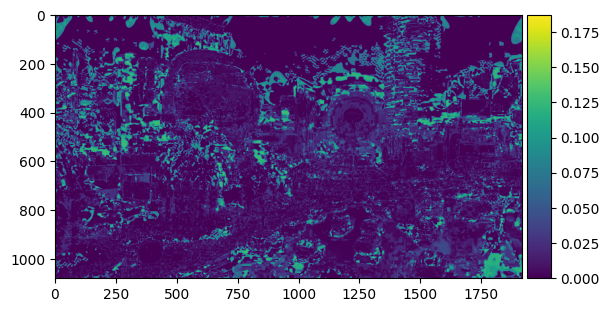}}
        \\
  \caption{Image difference maps for a pair of contrast-distorted images}
  \label{fig:contrast_diff} 
\end{figure}

\begin{figure}
    \centering
  \subfloat[Reference Image]{%
       \includegraphics[width=0.45\linewidth]{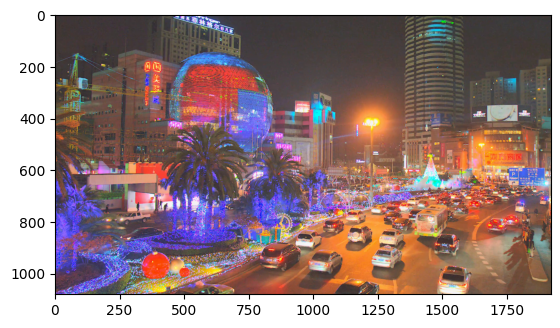}}
    \hfill
  \subfloat[Test Image]{%
        \includegraphics[width=0.45\linewidth]{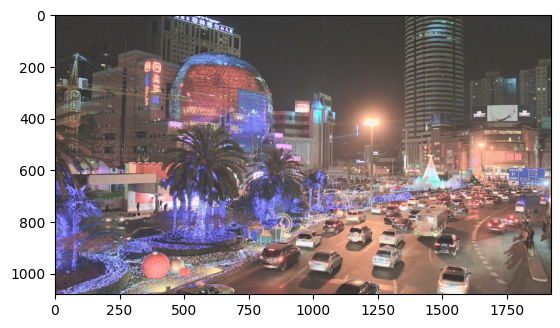}}
        \\
  \subfloat[Total Error (Baseline)]{%
       \includegraphics[width=0.45\linewidth]{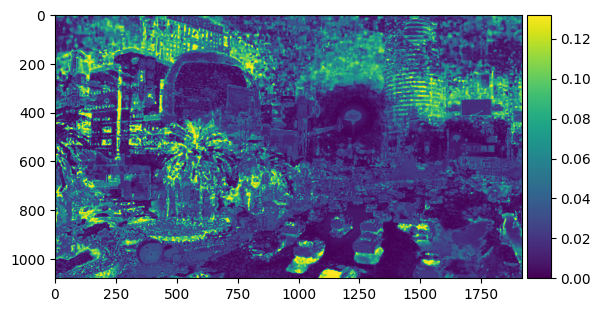}}
    \hfill
  \subfloat[Total Error (Edge-Aware)]{%
        \includegraphics[width=0.45\linewidth]{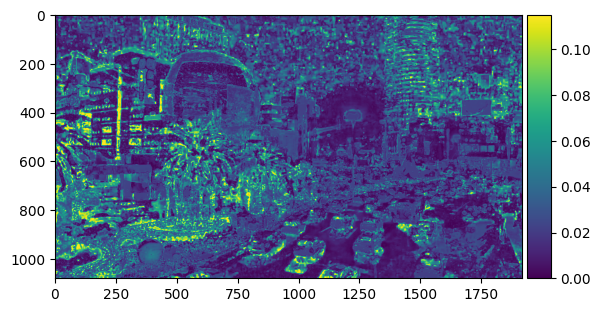}}
        \\
  \subfloat[Intensity Error (Baseline)]{%
       \includegraphics[width=0.45\linewidth]{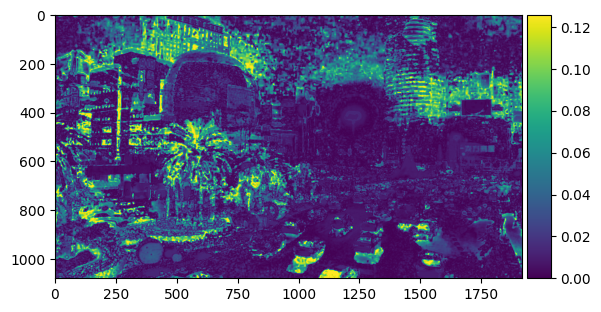}}
    \hfill
  \subfloat[Intensity Error (Edge-Aware)]{%
        \includegraphics[width=0.45\linewidth]{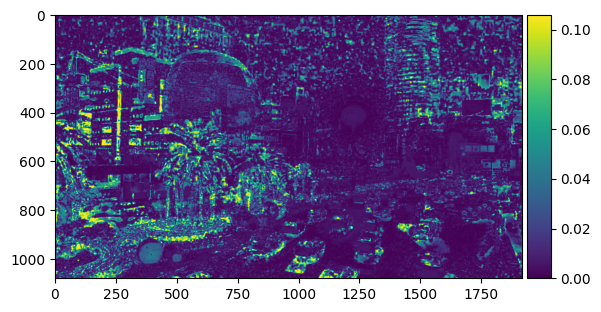}}
        \\
\subfloat[Chroma Error (Baseline)]{%
       \includegraphics[width=0.45\linewidth]{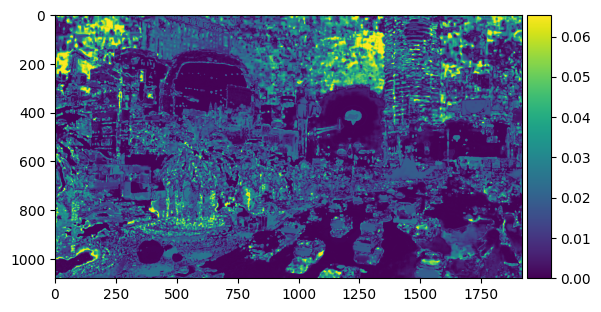}}
    \hfill
  \subfloat[Chroma Error (Edge-Aware)]{%
        \includegraphics[width=0.45\linewidth]{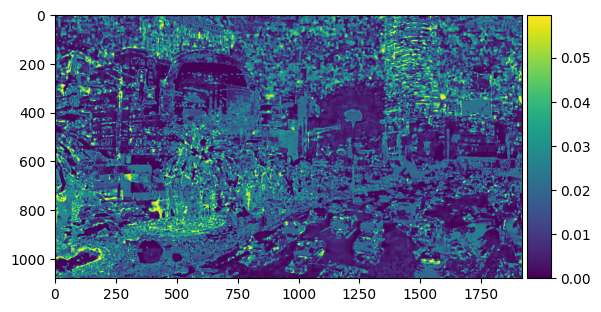}}
        \\
\subfloat[Hue Error (Baseline)]{%
       \includegraphics[width=0.45\linewidth]{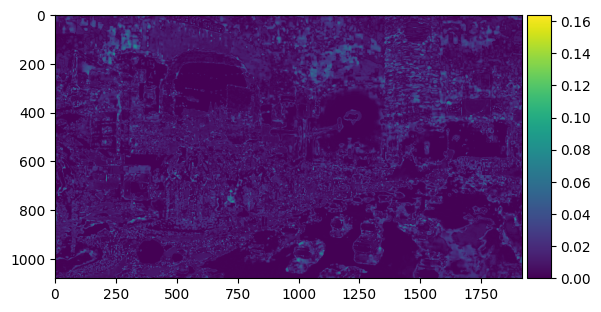}}
    \hfill
  \subfloat[Hue Error (Edge-Aware)]{%
        \includegraphics[width=0.45\linewidth]{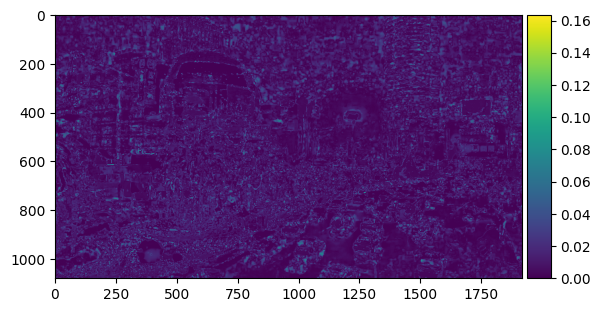}}
        \\
  \caption{Image difference maps for a pair of desaturation-distorted images}
  \label{fig:desat_diff} 
\end{figure}

The key advantage of the edge-aware model is expected to be in providing higher quality error maps that localize differences better. To demonstrate that we do indeed achieve this behavior, we present sample distortion maps from both distortion classes for the baseline and edge-aware models in Figures \ref{fig:contrast_diff} and \ref{fig:desat_diff}.

In Figure \ref{fig:contrast_diff}, we observe that the proposed edge-aware model greatly reduces the spurious CSF filtering-related artifacts present in the baseline model's output, such as the oblique patches emerging out of the buildings at the top of columns 750 and 1600 (approximately). Secondly, the baseline model predicts a significant difference around row 400 and column 1200, which corresponds to the lights on the side of the building. However, the actual difference is not so pronounced, and the edge-aware model rightly predicts that the main difference in the region is the halo of the light.

In Figure \ref{fig:desat_diff}, we again observe two main improvements that the edge-aware model makes over the baseline. Firstly, the baseline model overestimates the difference above the lights (row 400, column 1200) and above the dome (row 200, column 500). We attribute this to the CSF ``leakage'' from the lights nearby. The edge-aware model corrects this overestimation. Moreover, the edge-aware model provides a more uniform estimation of the error on the building around column 1500, and provides cleared outlines around errors corresponding to cars on the road.

In summary, our experiments demonstrate that the edge-aware iCAMDiff model provides more robust predictions of both local and aggregate image differences across two distortion types.

\section{Conclusion}
\label{sec:conclusion}
In this paper, we have reviewed color appearance, color difference, and contrast sensitivity models. We have motivated the need for image color appearance and color difference models that combine aspects of all the aforementioned models, and described two state-of-the-art models that attempt to tackle the problem. We then motivated and derived edge-aware image color appearance and difference models, and evaluated their performance for two distortion classes, demonstrating their utility.

While this work offers a promising case for edge-aware image difference models, there are still gaps that must be addressed. Firstly, a more robust evaluation must be conducted using a larger dataset, after obtaining subjective data specific to such distortions. In addition, a key drawback of our model is that the iterative edge-aware filtering method used is slow. This may be accelerated using better fast approximations to bilateral filters \cite{ref:fast_bilateral}, or simpler isotropic filtering techniques \cite{ref:eilertsen}. This work may also be extended to HDR images using the HDR UCSs and CSFs mentioned above.






\clearpage

\bibliographystyle{ieeetr}
\bibliography{refs}

\end{document}